\documentclass[10pt,letterpaper]{article}
\usepackage{graphicx}
\usepackage{epstopdf}
\begin{document}

\markboth{Reginald D. Smith}{An improved estimate of the inverse binary entropy function}

\title{An improved estimate of the inverse binary entropy function}

\author{Reginald D. Smith \\ Supreme Vinegar LLC \\ 3430 Progress Dr. Suite D, Bensalem, PA 19020 \\
rsmith@supremevinegar.com}
\date{May 5, 2020}

\maketitle

\begin{abstract}
Two estimates for the inverse binary entropy function are derived using the property of information entropy to estimate combinatorics of sequences as well as related formulas from population genetics for the effective number of alleles. The second estimate shows close correspondence to the actual value of the inverse binary entropy function and can be seen as a close approximation away from low values of binary entropy where $p$ or $1-p$ are small.
\end{abstract}

\section{The information entropy function}
\noindent While the concept of entropy first originated in thermodynamics with the German physicist Rudolf Clausius, it has reappeared and been reformulated in similar ways across a wide variety of disciplines in science and mathematics. Information entropy, first formulated by Claude Shannon in his seminal work on information theory \cite{shannon48}, was derived as an expression to calculate the capacity of a channel to transmit information given a discrete distribution of states or symbols. Where there are $M$ symbols each with probability $p_k$ for symbol, $k$, the information entropy, $H$, is defined as

\begin{equation}
H = -\sum_{k=1}^M -p_k \log p_k
\end{equation}

When the natural logarithm is used, the entropy is said to be measured in `nats' but $\log_2$ is often used and entropy is said to be measured in `bits' under this formulation. Where $M=2$, the entropy function is known as the binary entropy function with a simplified expression of

\begin{equation}
H = -p \log p - (1-p) \log (1-p)
\end{equation}

\subsection{Entropy and combinatorics}

Besides its nominal definition, entropy has a surprising and useful function for determining the expected number of sequence combinations length $L$ for a given distribution. In short, the expected number of sequences of length $L$, $N(L)$ for a distribution with entropy $H$ is calculated as \cite{shannonweaver59}

\begin{equation}
N(L) = e^{LH}
\label{combo}
\end{equation}

\subsection{The inverse binary entropy function}

While the binary entropy is easily calculated from a distribution defined by a Bernoulli trial, where the two possible outcomes occur with probabilities $p$ and $1-p$, calculating the inverse and finding the two values of $p$ and $1-p$ for a given value of entropy is a longstanding and unsolved problem. Despite the simple expression of the binary entropy equation, it is unclear if a closed form expression for the inverse, $H^{-1}(x)$ exists. Estimates have been given such as lower and upper bounds \cite{calabro09}. This paper will present two estimates for the inverse entropy function, with the second being most exact, derived using the combinatoric nature of entropy. The combinatoric approach is inspired by formulations in population genetics where the expected frequency of homozygous genotypes at a locus is used to estimate the expected number of alleles at that locus; a concept analogous to that calculated in equation \ref{combo}.

\subsection{Combinatorics and the genotype of a single locus}

The key concerns of population genetics are describing and explaining the nature of genetic variation within and between populations of the same species and understanding the forces that affect these across generations of inheritance. The typical starting unit of analysis is the locus, a unit of genetic inheritance which (in diploid organisms) has two alleles, one received from each parent, and is a discrete unit of inheritance whose basic laws were first described by Gregor Mendel. Each allele in a locus can come in any number of different types though often in theory and practice there are only two common variants for the allele whose frequencies are designated as $p$ and $q=1-p$. This is termed a bi-allelic locus.

There are many methods to measure the genetic diversity within a single population or between two populations at one or multiple loci. For a single bi-allelic locus, however, one of the simplest measures of genetic diversity is to analyze its expected homozygosity (both alleles being identical) and expected heterozygosity (both alleles being different). In bi-allelic loci, under Hardy-Weinberg equilibrium which assumes random mating and no selection on genotypes, the expected homozygosity for each allele variant are $p^2$ and $q^2$ and the expected heterozygosity is $2pq$.

Heterozygosity can be used as a proxy for diversity at a locus since its value increases as the frequency of both alleles approaches 1/2. However, another informative and more easily interpretable measure of locus diversity is the effective number of alleles.

\section{The effective number of alleles}

\subsection{Kimura-Crow formulation of effective number of \\ alleles}

First derived by famed population geneticists Motoo Kimura and James Crow \cite{kimuracrow64} the effective number of alleles, $N_a$ is simply the inverse of the expected homozygosity at a locus.

\begin{equation}
N_a = \frac{1}{p^2 + q^2} = \frac{1}{p^2 + (1-p)^2}
\label{effalleles}
\end{equation}

The effective number of alleles is a measure of diversity at a bi-allelic locus being a minimum 1 if one allele is fixed ($p$ or $q$ equals one) or a maximum of 2 when diversity is at a maximum and $p=q=1/2$.

\subsection{Inverse entropy estimate using Kimura-Crow \\ formulation}

The effective number of alleles can be viewed as a combinatoric measure analogous to $N$ for a sequence of length $L=1$ which is the effective number of symbols given the entropy of a distribution. Therefore, we can approximately equate $N(1)$ and $N_a$ using equations \ref{combo} and \ref{effalleles}.

\begin{equation}
e^H \approx \frac{1}{p^2 + (1-p)^2}
\end{equation}

This allows us to derive an estimate for $p$ as

\begin{equation}
H^{-1}(H) = p \approx \frac{1}{2} \pm \frac{1}{2} \sqrt{2e^{-H}-1}
\label{est1}
\end{equation}

\begin{figure}[h]
\begin{center}
\includegraphics[height=3in, width=3in]{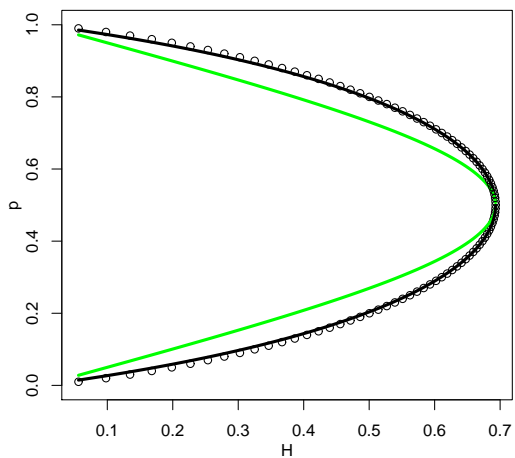}
\end{center}
\caption{A graph of both branches  of the inverse entropy function (points) along with the estimate provided for by equation \ref{est1} (green) and the revised derivation in equation \ref{est2} (black).}
\label{graphcompare}
\end{figure}

This is a rough approximation of both possible values of $p$ given the entropy. However, as shown in Figures \ref{graphcompare} and \ref{compare}, it is only a good estimate at the boundaries, where entropy is nearly zero, and near the maximum entropy value of $\log 2$.

\subsection{Improved estimate based on reduction from the \\ maximum possible effective number of symbols}

Given the range of the effective number of symbols for $L=1$ from one to two, another approach to estimate the effective number of symbols (alleles) is to start with the maximum of 2 and reduce based on changes in $p$ that reflect increased or decreased diversity. In short, increases in homozygosity reduce diversity. Therefore, a modified expected number of symbols can be expressed as

\begin{equation}
N(1) \approx 2-p^2 -(1-p)^2 + 2p(1-p)
\end{equation}

Using $N(1)=e^H$ we can then derive another estimate for inverse entropy

\begin{equation}
H^{-1}(H) = p \approx \frac{1}{2} \pm \frac{1}{2} \sqrt{2-e^H}
\label{est2}
\end{equation}

\begin{figure}[h]
\begin{center}
\includegraphics[height=3in, width=3in]{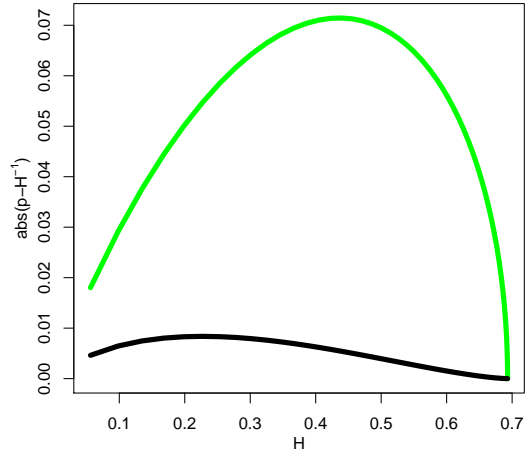}
\end{center}
\caption{Absolute value of error of estimates of $H^{-1}$ based on the effective number of alleles derivation from equation \ref{est1} (green) and the revised derivation in equation \ref{est2} (black).}
\label{compare}
\end{figure}

Figure \ref{compare}, shows this is a much superior approximation with a difference of less than 0.01 across all values and becoming nearly exact as entropy approaches $\log 2$. Granted, the overall precision is reduced at low values of $p$ or $1-p$ since the relative error is increased when entropy is small.

\section{Conclusion}

This paper presents improved estimates for the inverse binary entropy function. While neither is exact, the second estimate from equation \ref{est2} performs very well for values of entropy significantly different from zero and could feasibly be used in applications where approximations are acceptable.

\end{document}